\documentclass[11pt,a4paper]{revtex4-1}

\usepackage{amsfonts,amsmath,amssymb,xcolor}
\usepackage{amsthm,amscd,ulem,bm}
\usepackage[dvipdfmx]{graphicx}
\usepackage{slashed}
\usepackage{wrapfig}
\usepackage[truedimen, left=23truemm, right=23truemm, top=20truemm, bottom=10truemm,
 footskip=15truemm, includefoot]{geometry}

\newcommand{\nc}{\newcommand}
\nc{\obib}[2]{\frac{d {#1}}{d {#2}}}
\nc{\pbib}[2]{\frac{\partial {#1}}{\partial {#2}}}
\def\rot{\mbox{rot}}

\begin{document}

\title{
Generalized Mathisson-Papapetrou-Tulczyjew-Dixon Equations}
\author{Takayoshi Ootsuka}
\email{ootsuka@cosmos.phys.ocha.ac.jp}
\affiliation{Department of Physics, Meisei University, 2-1-1 Hodokubo Hino, Tokyo191-8506, Japan}
\author{Ryoko Yahagi}
\email{yahagi@rs.tus.ac.jp}
 \affiliation{Department of Physics, Faculty of Science, 
Tokyo University of Science, 1-3 Kagurazaka, Shinjuku-ku, Tokyo, Japan}

\date{\today}

%%%%%%%%%%%%%%%%
\begin{abstract}
%%%%%%%%%%%%%%%%

We derive two generalizations of Mathisson-Papapetrou-Tulczyjew-Dixon equations from Casalbuoni-Brink-Schwarz type pseudoclassical Lagrangians of Majorana spinors on a Riemann-Cartan spacetime. One has a ``color'' freedom, which makes the equations of motion also be a generalization of Wong equations. The other is a spinor model coupled with a Rarita-Schwinger field to preserve the supersymmetry. The coupling to the torsion is modified due to the existence of the Rarita-Schwinger field. In both extensions, the Tulczyjew condition is automatically satisfied.

%%%%%%%%%%%%%%
\end{abstract}
%%%%%%%%%%%%%%

\maketitle

%%%%%%%%%%%%%%%%%%%%%%
\section{Introduction}
%%%%%%%%%%%%%%%%%%%%%%

The equations of motion of an extended body on a curved spacetime were first considered by Mathisson in 1937 \cite{Mathisson}. Subsequently, other contributions were made by Papapetrou \cite{Papapetrou}, Tulczyjew \cite{Tulczyjew}, and Dixon \cite{Dixon}. We call the equations Mathisson-Papapetrou-Tulczyjew-Dixon (MPTD) equations in this paper. A good collection of works and applications concerning the MPTD equations is compiled in \cite{Puetzfeld}.

Lagrangian point of view for the equations has also been discussed in some literature \cite{Leclerc,Lompay,Unal,Ramirez}. Especially, Leclerc \cite{Leclerc} and \"{U}nal \cite{Unal} have considered a spinning particle on a Riemann-Cartan spacetime and appended a term depending on torsion to the original MPTD equations. Barducci {\it et al.} \cite{Barducci} have used a pseudoclassical approach to address the MPTD equations. A pseudoclassical model has Grassmann coordinates which are a vestige of a spinor quantum field in the semiclassical limit. One of the simplest models of a pseudoclassical particle is given by Cassalbuoni \cite{Casalbuoni} and Brink and Schwarz \cite{Brink-Schwarz}. The Lagrangian of this model comprises Majorana spinors which represents the spin degrees of freedom of the particle. In this paper, we extend this model, which is originally defined on the Minkowski spacetime, to the one on a Riemann-Cartan spacetime introducing the spin connection. Unlike \cite{Barducci}, our Lagrangian is in a form of simple square root. Two-dimensinal model has been discussed in our previous study \cite{OITY}.

In the following section, we introduce general notations used in this paper and the first model that derives the combination of MPTD and Wong equations. In section 3, the second model with the Rarita-Schwinger field is discussed. Note that we use left derivatives throughout this paper.

%%%%%%%%%%%%%%%%%%%%%%%%%%%%%%%
\section{MPTD-Wong equations} \label{mptd-wong}
%%%%%%%%%%%%%%%%%%%%%%%%%%%%%%%

%%%%%%%%%%%%%%%%%%%%%%%%%%%%%%%
\subsection{Model}
%%%%%%%%%%%%%%%%%%%%%%%%%%%%%%%

Our pseudoclassical Lagrangian is defined on a supermanifold, where both Grassmann odd and even coordinates are considered. The notation $x^\mu, \, \mu=0,1,2,3$ is used for the Grassmann even coordinates which describe the spacetime. For Grassmann odd coordinates, we use the real values $\xi^{A \alpha}, \, A=0,1,2,3, \, \alpha=1, \cdots, N$, where the index $\alpha$ represents an additional $N$-state internal degree of freedom, which we call ``color'' of the pseudoclassical particle. The $(4,4N)$-dimensional supermanifold concerned is denoted as $M^{(4,4N)}$. The submanifold $M^{(4,0)}$ is a Riemann-Cartan spacetime with a metric $g$ with signature $(+,-,-,-)$ and a spin connection $\displaystyle \omega = \frac14 \omega_{ab} \gamma^{ab} = \omega_c \theta^c = \frac14 \omega_{abc} \gamma^{ab} \theta^c, \, a,b,c \cdots=0,1,2,3$. We define $\displaystyle \gamma^{ab} = \frac12 \left(\gamma^a \gamma^b - \gamma^b \gamma^a \right)$ where $\gamma^a$ are gamma matrices, and $\theta^a=e^a {}_\mu(x) dx^\mu$ where $e^a {}_\mu(x)$ are vierbeins. 

We consider a pseudoclassical Lagrangian given by
\begin{align}
 L=-m\sqrt{\eta_{ab} \Pi^a \Pi^b}, 
 \quad
 \Pi^a=\theta^a + \langle \xi | \gamma^5 \gamma^a | D\xi \rangle,
\end{align}
where
\begin{align}
 | D\xi \rangle 
 = &
 |d\xi\rangle + \omega |\xi \rangle 
 + A |\xi \rangle
 + i \kappa \theta_a \gamma^a |\xi \rangle \notag \\
 = &
 e_\alpha \otimes | _A \rangle \left( d\xi^{A \alpha} + \frac{1}{4} \omega_{abc}
 (\gamma^{ab})^A {}_B \theta^c \xi^{B \alpha} 
 + A^\alpha{}_{\beta a} \theta^a \xi^{A \beta}
 + i \kappa \theta_a (\gamma^a)^A {}_B \xi^{B \alpha}
 \right).
\end{align}
Here, $m$ is the mass, the speed of light is set to $1$ and $\gamma^5=
 i \gamma^0 \gamma^1 \gamma^2 \gamma^3$. Note that this Lagrangian can be also considered as a Finsler metric on a supermanifold $M^{(4,4N)}$, which plays an important role to define the Lie derivative in the next subsection. By introducing the constant $\kappa$, we have the renormalized mass $m':= m (1 + i \kappa \langle \xi | \gamma^5 | \xi \rangle )$. Accordingly, we take the bare and renormalized momentum as
\begin{align}
 & P_a:= \pbib{L}{\Pi^a}
 = m^2 \frac{\Pi_a}{L}
 , \quad
  P_a P^a = m^2, \\
& \tilde{P}_a:= P_a (1 + i \kappa \langle \xi | \gamma^5 | \xi \rangle ) 
 = \frac{m'}{m} P_a , \quad
 \tilde{P}_a \tilde{P}^a = (m')^2.
\end{align}
If $\xi^{A \alpha} = 0$, it is the normal Lagrangian of a relativistic particle. $\xi^{A \alpha}$ are coefficients of a spinor $| \xi \rangle = e_\alpha \otimes |_A \rangle \xi^{A \alpha}$ with respect to the spinor basis $|_A \rangle$ and the basis $e_\alpha$ of a real representation space of the Lie algebra conserned. We assume the representation space has an inner product $\langle e_\alpha , e_\beta \rangle = h_{\alpha\beta}$ (similarly, $\langle e^\alpha , e^\beta \rangle = h^{\alpha\beta}$ and $\langle e_\alpha , e^\beta \rangle = \delta_\alpha^\beta$ for the dual basis $e^\alpha$). We also introduce a gauge field $A = A^i T_i$ which acts on a spinor $| \xi \rangle$ as
\begin{align}
 A |\xi \rangle
 = e_\alpha \otimes |_A \rangle A^\alpha{}_\beta \xi^{A \beta}
 = e_\alpha \otimes |_A \rangle A^\alpha{}_{\beta a} \theta^a \xi^{A \beta},
 \quad
 A^\alpha{}_\beta := A^i (T_i)^\alpha{}_\beta 
 = A^i {}_a \theta^a (T_i)^\alpha{}_\beta,
\end{align}
where $T_i$ form the basis of the Lie algebra for the model. In this representation, the basis $T_i$ satisfies $(T_i)_{\alpha\beta}=-(T_i)_{\beta\alpha}$. By introducing an anti-symmetric spinor metric $B_{AB}$ \cite{Chev, Suz} which has the properties
\begin{align}
 B \gamma^a B^{-1} = \, ^t \gamma^a, \quad 
 B_{AB}=-B_{BA}, \quad
 B^2=1,
\end{align}
the coefficients of the metric dual spinor are determined as $\xi_{A \alpha} = \xi^{B \beta} B_{BA} h_{\alpha\beta}$. The dual spinor is defined by $\langle \xi | = \xi_{A \alpha} e^\alpha \otimes \langle ^A |$, where $\langle ^A |$ is the dual basis of $|_A \rangle$. The inner product between spinors $| \xi \rangle$ and $| \xi' \rangle$ becomes
\begin{align}
 \langle \xi | \xi' \rangle = \xi_{A \alpha} \xi'^{A \alpha} = \xi^{A \alpha} B_{AB} h_{\alpha\beta} \xi'^{B \beta}= - \xi^{A \alpha} \xi'_{A \alpha}.
\end{align}
Due to the definition of the matrix $B_{AB}$, gamma matrices $(\gamma^a)_{AB}:={(\gamma^a)^C}_B B_{CA}$ admit the following properties:
\begin{align}
 & (\gamma^a)_{AB}=-(\gamma^a)_{BA}, 
 \hspace{10mm}
 (\gamma^5)_{AB}=-(\gamma^5)_{BA}, \notag \\
 & (\gamma^{ab})_{AB}:= 
 \frac12 \left(\gamma^a \gamma^b - \gamma^b \gamma^a \right)_{AB}
 = (\gamma^{ab})_{BA},
 \hspace{10mm}
 (\gamma^5 \gamma^a)_{AB}=(\gamma^5 \gamma^a)_{BA}, \notag \\
 & (\gamma^{abc})_{AB}:= 
 \frac16 \left(\gamma^a \gamma^b \gamma^c + \gamma^b \gamma^c \gamma^a
 + \gamma^c \gamma^a \gamma^b - \gamma^a \gamma^c \gamma^b
 - \gamma^c \gamma^b \gamma^a - \gamma^b \gamma^a \gamma^c \right)_{AB}
 = (\gamma^{abc})_{BA}, \notag \\
 & (\gamma^5 \gamma^{ab})_{AB}=(\gamma^5 \gamma^{ab})_{BA},
 \hspace{5mm}
 (\gamma^5 \gamma^{abc})_{AB}=-(\gamma^5 \gamma^{abc})_{BA}.
\end{align}
We also use the commutation relations
\begin{align}
 \left[\frac{\gamma^{ab}}{2}, \gamma^c\right]
 = & 
 \left(\eta^{bc} \gamma^a
 - \eta^{ac} \gamma^b\right) \label{ggg} \\
 \left[\frac{\gamma^{ab}}{2}, \frac{\gamma^{cd}}{2}\right]
 = &
 \left(
 \eta^{bc} \frac{\gamma^{ad}}{2}
 - \eta^{ac} \frac{\gamma^{bd}}{2}
 - \eta^{bd} \frac{\gamma^{ac}}{2}
 + \eta^{ad} \frac{\gamma^{bc}}{2}\right) \label{gggg}
\end{align}
in the rest of the discussion.

%%%%%%%%%%%%%%%%%%%%%%%%%%%%%%%
\subsection{Vector fields and the equations for their conjugate quantities}
%%%%%%%%%%%%%%%%%%%%%%%%%%%%%%%

Let $\displaystyle v=v^I \pbib{}{z^I}$ be a vector field on $M^{(4,4N)}$. Here, we set $(z^I) := (x^\mu, \xi^{A\alpha})$, where capital Roman letter $I$ stands for both spacetime and spinor indices. We define Lie derivative of the Lagrangian (Finsler metric) with respect to $v$ \cite{Killing} as
\begin{align}
 {\cal L}_v L
 := v^I \pbib{L}{z^I} + dv^I \pbib{L}{dz^I}
 = d \left( v^I \pbib{L}{dz^I} \right)
 + v^I \left\{ \pbib{L}{z^I} - d \left( \pbib{L}{dz^I} \right) \right\}.
\end{align}
The derivative $d$ here is a total derivative that is calculated as
\begin{align}
df(z,dz)=dz^I \pbib{f}{z^I} + d^2 z^I \pbib{f}{dz^I},
\end{align}
for a function $f(z,dz)$. 
When the Euler-Lagrange equations $\displaystyle \pbib{L}{z^I} - d \left( \pbib{L}{dz^I} \right) = 0$ are satisfied, this relation is equivalent to the identity
\begin{align}
 v^I \pbib{L}{z^I} + dv^I \pbib{L}{dz^I}
 = d \left( v^I \pbib{L}{dz^I} \right),
 \label{lie}
\end{align}
for any $v^I$. This is the evolution equation of the quantity $\displaystyle v^I \pbib{L}{dz^I}$. In the following discussion, the relation \eqref{lie} is repeatedly used with various vector fields. To specify the physical meaning of the quantity $\displaystyle v^I \pbib{L}{dz^I}$, we also mention the related symmetry beforehand.

%%%%%%%%%%%%%%%%%%%%%%%%%%%%%%%
\subsubsection{Color symmetry}
%%%%%%%%%%%%%%%%%%%%%%%%%%%%%%%

Firstly, we show the color symmetry. The gauge transformation is expressed as
\begin{align}
 \delta |\xi \rangle = \epsilon | \xi \rangle
 \quad \left( \delta \xi^{A \alpha} = \epsilon^\alpha{}_\beta \xi^{A \beta}
 \right)
 ,
 \hspace{5mm}
 \delta A = -D\epsilon = -\left( d \epsilon +  [ A, \epsilon] \right),
 \hspace{5mm}
 \epsilon^\alpha{}_\beta(x):=\epsilon^i(x) (T_i)^\alpha{}_\beta
\end{align}
where $\epsilon^i(x)$ are arbitrary functions of $x^\mu$ and $\epsilon_{\alpha \beta}= - \epsilon_{\beta \alpha}$. Accordingly, we have
\begin{align}
 \delta \langle \xi |
 = -\langle \xi | \epsilon
 ,
 \quad
 \delta | D \xi \rangle
 = \epsilon | D \xi \rangle,
\end{align}
which leads to $\delta \Pi^a = 0$.
Thus, the Lagrangian is invariant under this transformation.

The corresponding vector field is
\begin{align}
 v = \epsilon^\alpha{}_\beta \xi^{A \beta} \pbib{}{\xi^{A \alpha}},
\end{align}
where $\epsilon^\alpha{}_\beta$ are set to be antisymmetric constants from now on. Then, the equation \eqref{lie} becomes
\begin{align}
 \epsilon^\alpha{}_\beta \xi^{A \beta} 
 \pbib{L}{\xi^{A \alpha}}
 + \epsilon^\alpha{}_\beta d\xi^{A \beta} 
 \pbib{L}{d\xi^{A \alpha}}
 =
 d \biggl[
 \epsilon^\alpha{}_\beta \xi^{A \beta} \pbib{L}{d\xi^{A \alpha}} 
 \biggr],
\end{align}
i.e.,
\begin{align}
 - \frac12 \epsilon_{\alpha \beta} \left(
 A^\alpha {}_{\gamma} J^{\gamma \beta}
 + A^\beta {}_\gamma J^{\alpha \gamma}
 \right)
 =
 \frac12 \epsilon_{\alpha \beta} dJ^{\alpha \beta},
\end{align}
where we can define
\begin{align}
J^{\alpha \beta} := 2 P_a
 \langle \xi | \gamma^5 \gamma^a | _A^\alpha \rangle \xi^{A \beta} 
 = 2 P_a \xi_A^\alpha 
 ( \gamma^5 \gamma^a )^A {}_B \xi^{B \beta}
\end{align}
and identify it as a color corrent.
Eliminating $\epsilon_{\alpha \beta}$ from both sides, we have the covariant conservation of the color current $J^{\alpha \beta}$:
\begin{align}
 D J^{\alpha \beta} :=
 d J^{\alpha \beta} + A^\alpha {}_\gamma J^{\gamma \beta}
 + A^\beta {}_\gamma J^{\alpha \gamma}
 = 0.
 \label{dj}
\end{align} 
This is one of Wong equations. With the equation \eqref{dj}, it is easy to show
\begin{align}
 d \left( J_{\alpha \beta} J^{\alpha \beta} \right) = 0,
\end{align}
i.e., the norm of $J^{\alpha \beta}$ is conserved.

%%%%%%%%%%%%%%%%%%%%%%%%%%%%%%%
\subsubsection{Lorentz transformation}
%%%%%%%%%%%%%%%%%%%%%%%%%%%%%%%

Lorentz transformation is defined by
\begin{align}
 \delta \theta^a  = \epsilon^a {}_b \theta^b,
 \hspace{5mm}
 \delta |\xi \rangle = \frac14 \epsilon_{ab} \gamma^{ab} | \xi \rangle,
 \hspace{5mm}
 \delta \omega_{ab}
 = - d \epsilon_{ab} - \omega_{ac} \epsilon^c {}_b
 + \epsilon_{ac} \omega^c {}_b, \label{Lorentz}
\end{align}
where $\epsilon_{ab}(x)$ are antisymmetric functions of $x^\mu$.
Using identities \eqref{ggg} and \eqref{gggg}, we obtain
\begin{align}
 \delta \langle \xi |
 = -\frac14 \epsilon_{ab} \langle \xi | \gamma^{ab}, \quad
 \delta | D \xi \rangle
 = \frac14 \epsilon_{ab} \gamma^{ab} | D \xi \rangle.
\end{align}
Then, we have
\begin{align}
 \delta \biggl( \eta_{ab} \Pi^a \Pi^b \biggr)
 = 2 \eta_{ab} \Pi^a \delta \Pi^b
 = 2 \eta_{ab} \epsilon^b {}_c \Pi^a \Pi^c
 =0,
\end{align}
which shows the Lorentz invariance of the Lagrangian.

The vector field that corresponds to the Lorentz transformation is
\begin{align}
 v = \frac14 \epsilon_{ab} (\gamma^{ab})^A {}_B \xi^{B \alpha} 
 \pbib{}{\xi^{A \alpha}},
\end{align}
where $\epsilon_{ab}$ are now set to be constant for the following argument.
The equation \eqref{lie} for the above vector field becomes
\begin{align}
 \frac14 \epsilon_{ab} (\gamma^{ab})^A {}_B \xi^{B \alpha} 
 \pbib{L}{\xi^{A \alpha}}
 + \frac14 \epsilon_{ab} (\gamma^{ab})^A {}_B d\xi^{B \alpha} 
 \pbib{L}{d\xi^{A \alpha}}
 =
 d \biggl[
 \frac14 \epsilon_{ab} (\gamma^{ab})^A {}_B \xi^{B \alpha} \pbib{L}{d\xi^{A \alpha}}
 \biggr],
\end{align}
i.e.,
\begin{align}
 \epsilon_{ab} \biggl(
 P^a \langle \xi | \gamma^5 \gamma^b | D\xi \rangle
 -  \omega^a {}_c S^{bc}
 - i \kappa \theta^b P^a \langle \xi | \gamma^5 | \xi \rangle
 \biggr)
 & =
 - \frac12 \epsilon_{ab} dS^{ab},
\end{align}
where
\begin{align}
 S^{ab} :=- \frac12 P_c \langle \xi | \gamma^5 \gamma^{abc} | \xi \rangle
 \label{s}
\end{align}
is the spin of the particle. Defining the covariant derivative of the spin
\begin{align}
 DS^{ab} := d S^{ab} + \omega^a {}_c S^{cb} + \omega^b {}_c S^{ac},
\end{align}
we obtain
\begin{align}
 DS^{ab} 
 = \tilde{P}^a \theta^b - \tilde{P}^b \theta^a,
 \label{ds}
\end{align}
where an identity
\begin{align}
 \tilde{P}^a \theta^b - \tilde{P}^b \theta^a
 = - P^a \langle \xi |\gamma^5 \gamma^b |D\xi\rangle
 + P^b \langle \xi |\gamma^5 \gamma^a |D\xi\rangle
 + i \kappa \theta^b P^a \langle \xi | \gamma^5 | \xi \rangle
 - i \kappa \theta^a P^b \langle \xi | \gamma^5 | \xi \rangle
\end{align}
is used. The equation \eqref{ds} is the very one of the MPTD equations. The Tulczyjew condition
\begin{align}
 S^{ab} \tilde{P}_b 
 = 0
\end{align}
is automatically satisfied and the norm of the spin is conserved:
\begin{align}
 d \left( S_{ab} S^{ab} \right) = 0.
\end{align}

%%%%%%%%%%%%%%%%%%%%%%%%%%%%%%%
\subsubsection{Translation}
%%%%%%%%%%%%%%%%%%%%%%%%%%%%%%%

The vector field for the spacetime translation $\delta x^\mu = \epsilon^\mu$, where $\epsilon^\mu$ are constants, is defined as
\begin{align}
 v = \epsilon^\mu \pbib{}{x^\mu}. \label{trans}
\end{align}
The equation \eqref{lie} is
\begin{align}
 \epsilon^\mu \pbib{L}{x^\mu} + d \epsilon^\mu \pbib{L}{dx^\mu}  
 = 
 d \biggl[
 \epsilon^\mu \pbib{L}{dx^\mu} 
 \biggr],
\end{align}
i.e.,
\begin{align}
 \epsilon^\mu \biggl(
 \tilde{P}_a \partial_\mu e^a {}_\nu dx^\nu
 - \frac12 S^{bc} \partial_\mu \omega_{bc\nu} dx^\nu
 + \frac12 J^{\alpha \beta} \partial_\mu A_{\alpha\beta\nu} dx^\nu
 \biggr)
 =
 \epsilon^\mu 
 d \biggl(\tilde{P}_a e^a {}_\mu - \frac12 S^{bc} \omega_{bc\mu} 
 + \frac12 J^{\alpha \beta} A_{\alpha\beta\mu}
 \biggr),
\end{align}
and with the aid of the identity $\partial_\mu f_\nu dx^\nu - d f_\mu = i_{\partial_\mu} \rot (f_\nu dx^\nu)$ for a 1-form $f_\mu dx^\mu$ ($i_{\partial_\mu}$ stands for the interior product wirh $\partial_\mu$ and $\displaystyle \rot (f_\mu dx^\mu):= \frac12 \left(\pbib{f_\nu}{x^\mu}- \pbib{f_\mu}{x^\nu}\right)dx^\mu \wedge dx^\nu$ is used for the exterior derivative to differentiate it from the total derivative $d$), we obtain
\begin{align}
 d\tilde{P}_a e^a {}_\mu
 - \frac12 dS^{bc} \omega_{bc\mu}
 + \frac12 dJ^{\alpha \beta} A_{\alpha\beta\mu}
 =
 \tilde{P}_a i_{\partial_\mu} \rot \theta^a
 - \frac12 S^{bc} i_{\partial_\mu} \rot \omega_{bc}
 + \frac12 J^{\alpha\beta} i_{\partial_\mu} \rot A_{\alpha\beta}.
\end{align}
Substituting \eqref{dj}, \eqref{ds} and the definitions of torsion and curvature of the spacetime and the field strength of the gauge field $A$
\begin{align}
 & 
 \frac 12 T^a {}_{bc} \theta^b \wedge \theta^c
 :=
 \rot \theta^a + \omega^a {}_b \wedge \theta^b \\
 & 
 \frac12 R_{abcd} \theta^c \wedge \theta^d
 :=
 \rot \omega_{ab} + \omega_a {}^c \wedge \omega_{cb} \\
 & 
 \frac12 F_{\alpha\beta ab} \theta^a \wedge \theta^b
 :=
 \rot A_{\alpha \beta} 
 + A_\alpha {}^\gamma \wedge A_{\gamma\beta}
\end{align}
into the above equation, we finally obtain
\begin{align}
 D\tilde{P}_a 
 = 
 \tilde{P}_b T^b {}_{ac} \theta^c
 - \frac12 S^{bc} R_{bcad} \theta^d
 + \frac12 J^{\alpha \beta} F_{\alpha \beta a b} \theta^b. \label{DPTJ}
\end{align}
Here, the covariant derivative of the momentum is defined by
\begin{align}
 D \tilde{P}_a := d\tilde{P}_a - \omega^b {}_a \tilde{P}_b.
\end{align}
The second term in the equation \eqref{DPTJ} is seen in the MPTD equations and the first term with torsion appears because we are considering this model on a Riemann-Cartan spacetime, which is consistent with what  Leclerc \cite{Leclerc} and \"{U}nal \cite{Unal} have derived. The last term represents the field strength and the color current coupling in Wong equations. The equations
\eqref{dj}, \eqref{ds} and \eqref{DPTJ} express a generalization of the MPTD equations and Wong equations.

%%%%%%%%%%%%%%%%%%%%%%%%%%%%%%%
\section{MPTD equations with Rarita-Schwinger field}
%%%%%%%%%%%%%%%%%%%%%%%%%%%%%%%

%%%%%%%%%%%%%%%%%%%%%%%%%%%%%%%
\subsection{Model}
%%%%%%%%%%%%%%%%%%%%%%%%%%%%%%%

In this section, we consider a model which is invariant under simple supersymmetry. The parameter $\kappa$ and the gauge field $A$ are set to be zero. The spinor is now expressed as $| \xi \rangle = |_A \rangle \xi^A $ without color freedom. Instead, the Rarita-Schwinger field
\begin{align}
 | \psi \rangle = | \psi_\mu(x) \rangle dx^\mu
 = | _A \rangle \psi^A(x)
 = | _A \rangle \psi^A {}_\mu (x) dx^\mu
\end{align}
is introduced to preserve the supersymmetry. We consider the following Lagrangian throughout this section:
\begin{align}
 L=-m\sqrt{\eta_{ab} \Pi^a \Pi^b}, 
 \quad
 \Pi^a=\theta^a + \langle \xi | \gamma^5 \gamma^a | D\xi \rangle
 + \langle \xi | \gamma^5 \gamma^a | \psi \rangle, \quad
 | D\xi \rangle 
 = 
 |d\xi\rangle + \omega |\xi \rangle. \label{Lag_psi}
\end{align}
As in section \ref{mptd-wong}, several transformations and the conjugate equations are considered so that the generalized MPTD equations with the Rarita-Schwinger field.

%%%%%%%%%%%%%%%%%%%%%%%%%%%%%%%
\subsection{Vector fields and the equations for their conjugate quantities}
%%%%%%%%%%%%%%%%%%%%%%%%%%%%%%%

%%%%%%%%%%%%%%%%%%%%%%%%%%%%%%%
\subsubsection{Supersymmetry}
%%%%%%%%%%%%%%%%%%%%%%%%%%%%%%%

The model \eqref{Lag_psi} is invariant under the super transformation
\begin{align}
 \delta \xi^{A } = \epsilon^{A}, \quad
 \delta \theta^a  
 = 
 - \langle \epsilon |\gamma^5 \gamma^a | D \xi \rangle
 - \langle \epsilon |\gamma^5 \gamma^a | \psi \rangle, \quad
 \delta | \psi \rangle
 = - | D \epsilon \rangle
 = - \left( |d\epsilon\rangle + \omega |\epsilon \rangle
 \right),
\end{align}
where $\epsilon^{A}(x) $ are arbitrary Grassmann odd functions of $x^\mu$. In fact, this transformation directly leads to $\delta \Pi^a=0$. Unlike the original Casalbuoni-Brink-Schwarz Lagrangian defined on a flat spacetime, our superparticle model is supersymmetric even on a Riemann-Cartan spacetime.

Correspondingly, we consider a vector field
\begin{align}
 v= \epsilon^{A} \pbib{}{\xi^{A}}
\end{align}
for arbitrary Grassmann odd constants $\epsilon^A$. The equation \eqref{lie} becomes
\begin{align}
 \epsilon^{A} \pbib{L}{\xi^{A}} 
=
d \left[ \epsilon^{A} \pbib{L}{d\xi^{A}}
\right],
\end{align}
i.e.,
\begin{align}
 \epsilon^{A} P_a \biggl(
 - \langle _{A} | \gamma^5 \gamma^a | d\xi \rangle
 - \langle _{A} | \gamma^5 \gamma^a \omega | \xi \rangle
 - \langle _{A} | \gamma^5 \gamma^a | \psi \rangle
 - \langle \xi | \gamma^5 \gamma^a \omega | _{A} \rangle
 \biggr)
= 
\epsilon^{A} d Q_{A}.
\end{align}
Here, we take
\begin{align}
 Q_{A}  := - P_a \langle \xi | \gamma^5 \gamma^a | _{A} \rangle
\end{align}
as the supercharge. Defining the covariant derivative of the supercharge by
\begin{align}
 D Q_{A} 
 = 
 d Q_{A} - \frac{1}{4} \omega_{ab} Q_{B} (\gamma^{ab})^B{}_A
 = 
 \langle d Q | _{A } \rangle 
 - \langle Q | \omega | _{A} \rangle ,
\end{align}
we obtain
\begin{align}
 (D Q)_{A } 
 = - P_a \left(
 \langle _{A } | \gamma^5 \gamma^a | D\xi \rangle
 + \langle _{A } | \gamma^5 \gamma^a | \psi \rangle
 \right).
 \label{dq}
\end{align}

%%%%%%%%%%%%%%%%%%%%%%%%%%%%%%%
\subsubsection{Lorentz transformation}
%%%%%%%%%%%%%%%%%%%%%%%%%%%%%%%

In addition to the transformation \eqref{Lorentz}, we consider 
\begin{align}
 \delta |\psi \rangle = \frac14 \epsilon_{ab} \gamma^{ab} | \psi \rangle
\end{align}
for the Rarita-Schwinger field. This Lagrangian is also invariant under the Lorentz transformation. The equation \eqref{lie} for the vector field 
\begin{align}
 v = \frac14 \epsilon_{ab} (\gamma^{ab})^A {}_B \xi^{B} 
 \pbib{}{\xi^{A}}
\end{align}
with antisymmetric constant $\epsilon_{ab}$ becomes
\begin{align}
 \frac14 \epsilon_{ab} (\gamma^{ab})^A {}_B \xi^{B} 
 \pbib{L}{\xi^{A}}
 + \frac14 \epsilon_{ab} (\gamma^{ab})^A {}_B d\xi^{B} 
 \pbib{L}{d\xi^{A}}
 = 
 d \left[
 \frac14 \epsilon_{ab} (\gamma^{ab})^A {}_B \xi^{B } \pbib{L}{d\xi^{A }}
 \right],
\end{align}
i.e.,
\begin{align}
 \epsilon_{ab} \biggl(
 P^a \langle \xi | \gamma^5 \gamma^b | D\xi \rangle
 -  \omega^a {}_c S^{bc}
 - \frac14 P_c \langle \xi | \gamma^5 \gamma^{cab} | \psi \rangle
 + \frac12 P^a \langle \xi | \gamma^5 \gamma^b | \psi \rangle
 \biggr)
 = 
 - \frac12 \epsilon_{ab} dS^{ab}.
\end{align}
With the use of identities
\begin{align}
 P^a \theta^b - P^b \theta^a
 = 
 -P^a \langle \xi |\gamma^5 \gamma^b |D\xi\rangle
 + P^b \langle \xi |\gamma^5 \gamma^a |D\xi\rangle
 -P^a \langle \xi |\gamma^5 \gamma^b |\psi\rangle
 + P^b \langle \xi |\gamma^5 \gamma^a |\psi\rangle
\end{align}
and
\begin{align}
 \langle Q | \gamma^{ab} | \psi \rangle
 = 
 - P_c
 \langle \xi | \gamma^5 \gamma^{cab} | \psi \rangle
 - P^a
 \langle \xi | \gamma^5 \gamma^b | \psi \rangle
 + P^b
 \langle \xi | \gamma^5 \gamma^a | \psi \rangle,
\end{align}
we have
\begin{align}
 DS^{ab} 
 = P^a \theta^b - P^b \theta^a
 - \frac12 \langle Q | \gamma^{ab} | \psi \rangle.
 \label{dsplus}
\end{align}
This can be a generalized MPTD equation with the additional Rarita-Schwinger term coupled with the supercharge. The Tulczyjew condition 
\begin{align}
 S^{ab} P_b 
 = 0
\end{align}
is also satisfied.

%%%%%%%%%%%%%%%%%%%%%%%%%%%%%%%
\subsubsection{Translation}
%%%%%%%%%%%%%%%%%%%%%%%%%%%%%%%

For translation \eqref{trans}, the equation \eqref{lie} with respect to this Lagrangian is
\begin{align}
 \epsilon^\mu \pbib{L}{x^\mu} + d \epsilon^\mu \pbib{L}{dx^\mu}  
 = 
 d \biggl[
 \epsilon^\mu \pbib{L}{dx^\mu} 
 \biggr],
\end{align}
i.e.,
\begin{align}
\epsilon^\mu \biggl(
 P_a \partial_\mu e^a {}_\nu dx^\nu
 - \frac12 S^{bc} \partial_\mu \omega_{bc\nu} dx^\nu
 - \langle Q | \partial_\mu \psi_\nu \rangle dx^\nu \biggr) 
 =
 \biggl(P_a e^a {}_\mu - \frac12 S^{bc} \omega_{bc\mu} 
 - \langle Q | \psi_\mu \rangle
 \biggr).
\end{align}
Eliminating $\epsilon^\mu$, we have
\begin{align}
 dP_a e^a {}_\mu
 - \frac12 dS^{bc} \omega_{bc\mu}
 - \langle dQ | \psi_\mu \rangle 
 = 
 P_a i_{\partial_\mu} \rot \theta^a
 - \frac12 S^{bc} i_{\partial_\mu} \rot \omega_{bc}
 - \langle Q | i_{\partial_\mu} \rot \psi \rangle.
\end{align}
Substituting \eqref{dq} and \eqref{dsplus}, we obtain the other half of our generalized MPTD equations
\begin{align}
 DP_a 
 = &
 P_b \tilde{T}^b {}_a
 - \frac12 S^{bc} R_{bcad} \theta^d
 - \langle Q | i_{\partial_{e_a}} \rot \psi \rangle, \label{DP}
\end{align}
where we use the definition
\begin{align}
 D P_a := dP_a - \omega^b {}_a P_b,
\end{align}
and
\begin{align}
 \tilde{T}^b {}_a 
 :=
 T^b {}_{ac} \theta^c
 - \langle \psi_a | \gamma^5 \gamma^b | D\xi \rangle
 - \langle \psi_a | \gamma^5 \gamma^b | \psi \rangle. \label{tor}
\end{align}
This is the torsion modified by the Rarita-Schwinger field. If we consider the limit $\xi^A \to 0$, this modified torsion has the same expression as the one that appears in the context of supergravity.

%%%%%%%%%%%%%%%%%%%%%%%%%%%%%%%
\section{Discussion}
%%%%%%%%%%%%%%%%%%%%%%%%%%%%%%%

In this paper, we have derived the MPTD equations with torsion from a pseudoclassical Lagrangian of Majorana spinors on a supermanifold. Its Grassmann even part forms a Riemann-Cartan spacetime, which induces the torsion term to the original MPTD equations. We have also considered two types of generalization. The first one has additional internal degrees of freedom and renormalized mass. This adds the color current term that appeared in the Wong equations to our equations. For the other generalized model, we introduce the Rarita-Schwinger field to retain supersymmetry. This modifies the torsion consistently with supergravity. 

Since our second model is a relatively straightforward supersymmetric extension of the particle Lagrangian of general relativity, it can be inherently related to supergravity. Its geometric quantities such as connection and curvature on a supermanifold should be calculated for more concrete analysis. They will be treated in terms of Finsler geometry.

%%%%%%%%%%%%%%%%%%%%%%%%%%%

%%%%%%%%%%%%%%%%%%%%%

\end{document}